\documentclass[
  showpacs
  twoside,
  twocolumn,
  aps
  prlsdsB]{revtex4}

\usepackage[ngerman,american]{babel}
\usepackage{graphicx}
\usepackage{amsmath}
\usepackage{pxfonts}
\usepackage{wasysym}
\usepackage{manfnt}
\usepackage{color}
\usepackage{dingbat}
\usepackage{multirow}

\usepackage{epstopdf}

\begin{document}

\title{No-enclave percolation corresponds to holes in the cluster backbone}
\author{Hao Hu$^1$}
\author{Robert M. Ziff$^2$}
\author{Youjin Deng$^1$}
\email[]{yjdeng@ustc.edu.cn}
\affiliation{$^1$ Hefei National Laboratory for Physical Sciences at Microscale, Department of Modern Physics, University of Science
and Technology of China, Hefei 230027, China}
\affiliation{$^2$ Center for the Study of Complex Systems and Department of Chemical Engineering, University of Michigan, Ann Arbor, Michigan 48109-2136, USA}

\begin{abstract}
The no-enclave percolation (NEP) model introduced recently by Sheinman et al.\ can be mapped to a problem of holes within a standard percolation backbone, and numerical measurements of these holes gives the size-distribution exponent $\tau = 1.82(1)$ of the NEP model.  An argument is given that $\tau=1 + d_B/2 \approx 1.822$ where $d_B$ is the backbone dimension.  On the other hand, a model of simple holes within a percolation cluster implies $\tau = 1 + d_f/2 = 187/96 \approx 1.948$, where $d_f$ is the fractal dimension of the cluster, and this value is consistent with Sheinman et al.'s experimental results of gel collapse which gives $\tau = 1.91(6)$.  Both models yield a discontinuous maximum hole size at $p_c$, signifying explosive percolation behavior.   At $p_c$, the largest hole fills exactly half the system, due to symmetry.  Extensive numerical simulations confirm our results. 
\end{abstract}
\maketitle

Recently, Sheinman et al.\ \cite{SheinmanEtAl15} introduced the ÒNo-Enclave PercolationÓ  (NEP) model to explain the motor-driven collapse of a model cytoskeletal system studied by Alvarado et al.\ \cite{AlvaradoEtAl13}.  The NEP model, which has received a great deal of attention \cite{BenIssacFodorViscoWijlandGov15,JedrzejewskiChmielSznajdWeron15,SheinmanSharmaAlvaradoKoenderinkMacKintosh15b,DasLookmanBandi15,BoettcherWolleyMezaGolesHelbingHerrmann16,SegallTeomyShokef15,LeePruessner16,BarMajumdarSchehrMukamel16,KovacsJuhaszIgloi16,PruessnerLee16,SheinmanSharmaMacKintosh16},
 is based upon regular random percolation in which all clusters collapse, and what remains are solid clusters
that represent the gelled regions.
The authors of  \cite{SheinmanEtAl15} consider a region surrounded by sites of other clusters and the NEP clusters are composed of occupied sites within the region.  Reversing occupied and vacant sites or bonds, the problem can be thought of as finding the distribution of hole sizes within a single large cluster.  
Beside serving a successful model for active gels, the NEP is found to have a size-distribution exponent  (the so-called Fisher exponent) $\tau = 1.82 (1)$, which is less than the conventional lower bound value 2, and thus represents a distinct universality class from the standard random percolation.
However, the holes they study are not simple percolation holes, but, as we shall see, holes within the surrounding backbone.  In this Letter we derive universal expressions for the scaling behavior for both holes in the backbone, and  simple holes in percolation cluster.  We find that the latter gives scaling consistent with the experimental results of Ref. \cite{SheinmanEtAl15}, suggesting this is the appropriate model for their system.   Note that these holes are first step of the hierarchy of the connections of holes in a fully percolated system studied in  \cite{HuberKlebanZiff16}.

After this work was complete, a Comment \cite{PruessnerLee16} and Reply \cite{SheinmanSharmaMacKintosh16} were published discussing the admissibility of having a Fisher exponent $\tau$ less than 2, as the authors of Ref. \cite{SheinmanEtAl15} found.  Here we show that such exponents are entirely possible for systems with a cutoff, in agreement with \cite{SheinmanSharmaMacKintosh16}, and we verify this behavior with extensive simulations.

\begin{figure}[htbp] 
   \centering
   \includegraphics[width=2.5 in]{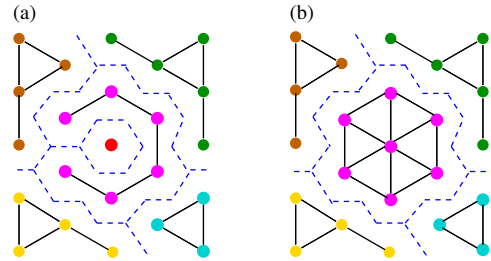} 
   \caption{ (color online.)  Diagram illustrating the NEP procedure of \cite{SheinmanEtAl15}.  (a) b-TR clusters and the dual honeycomb lattice (dashed bonds).   (b)  All components within a boundary of neighboring cluster sites become a no-enclave cluster, and remaining dashed bonds give the dual-lattice backbone.}
   \label{fig:bTR-backbone_color}
\end{figure}

In Ref. \cite{SheinmanEtAl15} (see supplementary material), Sheinman et al.\ consider clusters created by bond percolation on the triangular lattice (b-TR) at the threshold $p_c = 2 \sin \pi/18$ \cite{SykesEssam64}, and identify the external boundary by the sites of other clusters bordering the bond cluster.  Then they combine every site within the boundary into the no-enclave cluster.   Because they use the external sites to define the boundary, they also include openings of separation one lattice spacing in their clusters, as shown in Fig.\ \ref{fig:bTR-backbone_color}.  Closing these openings means that the boundary of the cluster becomes the external accessible hull \cite{AharonyGrossman86}, which has a fractal dimension of $4/3$. All remaining dual-lattice bonds (dashed lines in Fig.\ \ref{fig:bTR-backbone_color}(b)) are bi-connected, and accordingly they form the backbone of the dual-lattice cluster \cite{XuWangZhouGaroniDeng14}.  Thus, the no-enclave clusters are effectively holes within the backbone of the largest dual-lattice cluster.

Sheinman et al.\ find that their model yields a size-distribution exponent of $\tau = 1.82(1)$, while their experiments yield $\tau = 1.91(6)$.  Thus, their model supports the experimental result that $\tau < 2$, although the value of $\tau$ their model gives is somewhat low.

In this Letter we consider simple holes as well as the backbone holes.
We carried out extensive simulations of both for site percolation on the square (s-SQ) and triangular lattices (s-TR), and bond percolation on the square lattice (b-SQ).  We considered a square $L \times L$ system for the square lattice, and a rhombic $L \times L$ system for the triangular lattice.  
Periodic boundary conditions were implemented.
For the holes, we occupied the system with sites or bonds with probability $p$, identified largest black cluster, removed all remaining sites or bonds, then identified the white clusters.  For the backbone holes (bond percolation only), we identified the backbone first and then the holes within it.  We considered systems with $L = 8, 16, \ldots, 16384$ and 
carried out from $4 \times 10^5$ ($L = 16384$) to $5\times 10^6 $ ($L=8$) runs.  Note that we have not found previous work directly studying holes in percolation clusters.  The related problem of lacunarity in percolation clusters has been studied \cite{MandelbrotStauffer94,HoviAharonyStaufferMandelbrot96}.

\begin{figure}[htbp] 
   \centering
   \includegraphics[width=3.4in]{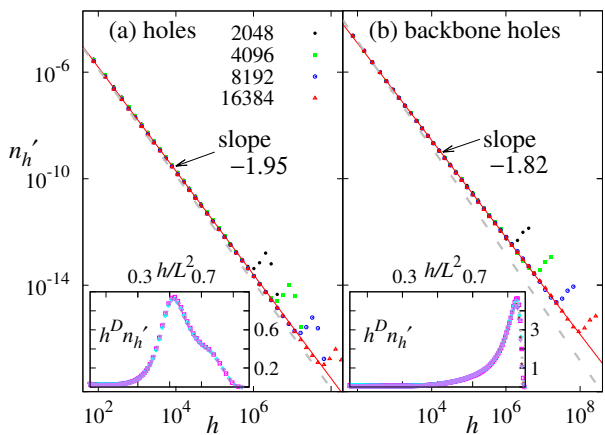}                                                                                                                                                                                                                                                                                                                                                                                                                                                                                                                                                                                                                                                                                                                                                                                                                                                                                                                                                                                                                                                                                                                                                                                                                                                                                                                                                                                                                                                                                                                                         
   \caption{(color online.) Scaled number of holes $n_h'=n_h(L,p_c) L^{d-D}$ (a) or backbone holes (b) of size $h$ as a function of $h$, showing respectively $\tau_h = 1.949(3)$ consistent with the prediction $187/96$ of  Eq.\ (\ref{tauholes}) and $\tau_b = 1.82(1)$ consistent with the prediction $1.822$ of Eq.\ (\ref{tauenclaves}), for systems of difference size $L$.  The inset shows $h^D n_h'$ vs.\ $h/L^2$ where $D$ is $d_f$ (a) or $d_B$ (b), demonstrating the accumulation in the distribution due to the largest clusters.  For (a) we used s-TR and for (b) we used b-SQ.  The dashed grey lines have a slope of $-2$ for comparison.}
   \label{fig:HoleSizeDist}
\end{figure}

The results for the size distribution are shown in Fig.\ \ref{fig:HoleSizeDist}.  For the backbone holes, we find $\tau_b = 1.82(1)$ which agrees with Sheinman et al.'s simulation.  For holes, we find a value of $\tau_h =  1.949(3)$  which is consistent with Sheinman et al.'s experimental results.  Thus, we argue that their experiment is more accurately modeled by the simple hole process.

\begin{figure}[htbp] 
   \centering
   \includegraphics[width=3.4in]{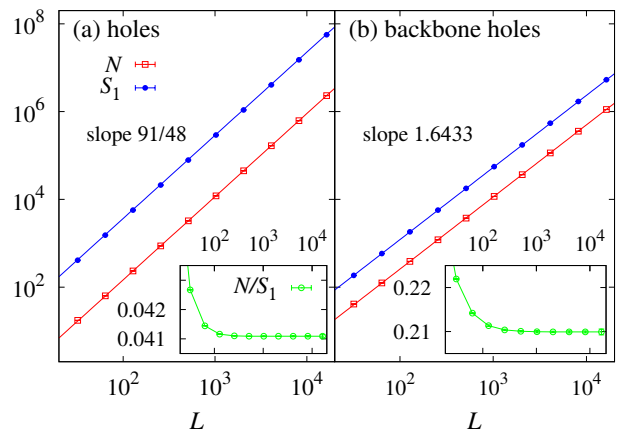} 
   \caption{(color online.) Log-log plot of the total number $N$ of holes (a) or backbone holes (b) in the largest cluster, and also the size $S_1$ of the largest cluster (a) or backbone (b), as a function of the size of the system $L$, with a slope equal to $d_f = 91/48$ for holes in a percolation cluster (a), and to $d_B = 1.6433$ for backbone holes  (b), showing that the number of cluster or backbone holes is proportional to the mass of the cluster (a) or backbone (b).  For (a) we use s-TR and for (b) we use b-SQ.  Inset shows $N/S_1$ which rapidly goes to a constant $\approx 0.0410907$ for holes and 0.209908 for backbone holes for large $L$.}
   \label{fig:total_num_holes}
\end{figure}

To derive a scaling relation for the holes within the percolation cluster, consider the hole-size distribution $n_h(L,p)$ equal to the number of holes per lattice site containing $h$ vertices in the largest cluster of an $L \times L$ system at bond occupation $p$.  At the critical point $p_c$, the total number of holes $N$ is proportional to the number of sites $s$ in the cluster, which scales as the system size as $\sim L^{d_f}$ where $d_f = 91/48$ is the cluster fractal dimension \cite{HuberKlebanZiff16}. This is demonstrated in Fig.\ \ref{fig:total_num_holes}(a) for s-TR, and we have also verified that proportionality for b-SQ.

Then it follows that $n_h(L,p)$  scales as
\begin{equation}
n_h(L,p) \sim L^{d_f - d} h^{-\tau_h} f_h(h/L^d)
\label{nh}
\end{equation}
where the scaling function $f_h(z)$ cuts off when $h$ is of order the size of the largest hole, which is proportional to $L^d$, where $d$ is the dimensionality which is always 2 in this paper.  The term at the beginning $L^{d_f - d}$ differs from the usual scaling of cluster size and reflects the fact that the density of holes of a given size $h$ goes to zero as the system size increases.
The form of Eq.(\ref{nh}) is identical to that proposed in \cite{SheinmanSharmaMacKintosh16}, noting that here $n_h$ represents the number of hole sites per lattice site, while in \cite{SheinmanSharmaMacKintosh16}, $n_s$ represents the total number of hole sites, so they differ by a factor of $L^d = M$.  Here we go on to find an exact expression for $\tau_h$, which was not found in Ref.\ \cite{SheinmanSharmaMacKintosh16}.

Next we need to make some considerations for the case that $\tau_h < 2$.  Normally, $\tau$ has to be greater than 2 so that the size distribution is normalizable: $\sum_{s=1}^\infty  s n_s \sim \sum^\infty s^{1-\tau} < \infty$.  However, for systems such as these where there is an upper cutoff to the sum and an asymptotically vanishing number of holes per lattice site, 
$\tau$ can be less than 2.  Say $n_h(L, p_c) \sim A h^{-\tau_h}$ with a cutoff $h_\mathrm{max}$.   Then
\begin{equation}
\sum_{h = 1}^{h_\mathrm{max}} h n_h(L, p_c) \sim \sum_{h = 1}^{h_\mathrm{max}} A h^{1 - \tau_h} +h_\mathrm{big}  \sim \frac{A}{2 - \tau_h} h_\mathrm{max}^{2 - \tau_h} + h_\mathrm{big} 
\label{hnh}
\end{equation}
This can remain constant as $h_\mathrm{max} \to \infty$ if $A \to 0$ with $A \sim h_\mathrm{max}^{\tau_h-2} \sim L^{d (\tau_h-2)}$.  Notice $\tau_h$ must be less than 2 for this to be possible.  We also split off $ h_\mathrm{big} $ which represents the largest hole, to allow a macroscopic occupancy of that quantity.  Comparing the scaling of $A$ with that of the leading term of (\ref{nh}), we have $d(\tau_h-2) = d_f - d$, which yields
\begin{equation}
\tau_h = 1 + \frac{d_f}{d} = \frac{187}{96} \approx 1.948
\label{tauholes}
\end{equation}
This value agrees with our simulation results $\tau_h = 1.949(3)$, and is  close to the experimental $\tau = 1.91(6)$ found in Ref.\ \cite{SheinmanEtAl15}.

The scaling relation (\ref{tauholes}) above is in the form of Mandelbrot's hyperscaling relation 
\begin{equation}
\tau = 1 + \frac{d_\mathrm{all}}{d_\mathrm{object}}
\label{mandelbrot}
\end{equation}
 (see \cite{Voss84}), where the objects combine together to make the ``all."  For example, in normal two-dimensional percolation where the ensemble of fractal clusters fill the non-fractal space, we have $d_\mathrm{all} = 2$ and $d_\mathrm{object} = 91/48$, yielding $\tau = 187/91 \approx 2.055$.  From our analysis we can see that for holes, $d_\mathrm{all}$ corresponds to the fractal dimension of the largest cluster.  It is also the union of the hulls of all the holes, which forms the fractal cluster.

\begin{figure}[htbp] 
   \centering
   \includegraphics[width=3.4in]{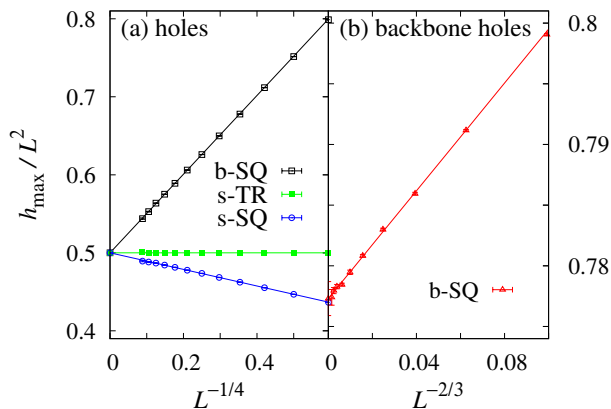} 
   \caption{(color online.) (a) The largest hole size $h_\mathrm{max}$ divided by the number of sites $L^2$, as a function of $L^{-1/4}$ for various systems, showing that this quantity is exactly 1/2 for s-TR, but only approaches 1/2 as $L \to \infty$ for s-SQ and b-SQ.  For b-SQ we measure the size of the hole as the number of sites it contains; if we measure the holes by vacant bonds, then $h_\mathrm{max}/L^2 =1/2$ for all $L$.  (b)  The fraction of the system filled by the largest backbone hole as a function of $L^{-2/3}$ for b-SQ.  Here, the fraction approaches 0.7772(4) as $L \to \infty$.  The scaling in both cases is $L^{d_H - d}$ where the hull dimension $d_H = 7/4$ (holes) and $4/3$ (backbones) because for asymmetric systems the hull contributes to the size of the largest hole $h_\mathrm{max}$ proportional to its length.}
  \label{fig:C1}
\end{figure}

In the case of the backbone holes, where the backbone has a dimension $d_{B} = 1.64336(10)$ \cite{DengBloeteNienhuis04,ZhouYangDengZiff12}, a similar argument gives
\begin{equation}
\tau_b = 1 + \frac{d_B}{2} = 1.82168(5)
\label{tauenclaves}
\end{equation}
which agrees with our measurements $\tau_b = 1.82(1)$, and also with Sheinman et al.'s numerical results, supporting the idea that the no-enclave clusters they consider are effectively dual-lattice backbone holes.

In Eq.\ (\ref{hnh}) we allowed for a large hole.  If fact, we measured the size of the largest hole in a periodic system of size $L \times L$.  On the torus, the largest hole can be a co-wrapping cluster on the torus, a non-wrapping cluster, or a cross-configuration wrapping in both directions (in which case the hole is external to the cluster).   For s-TR we found the average size of the largest hole to be exactly half the size of the system, while for the other systems we studied it approaches 1/2 as $L \to \infty$, as shown in Fig.\ \ref{fig:C1}.  To explain this, observe that the largest white cluster (hole) and the black cluster with every hole but the largest filled in are identical on the triangular lattice where $p_c = 1/2$ --- switching black and white does not change anything --- so on the average each is 1/2 the lattice.  For other lattices where there is not perfect self-duality or self-matching, one would expect the argument to hold only asymptotically for large $L$ at criticality.  This is indeed what we see.

\begin{figure}[htbp] 
   \centering
   \includegraphics[width=3.4in]{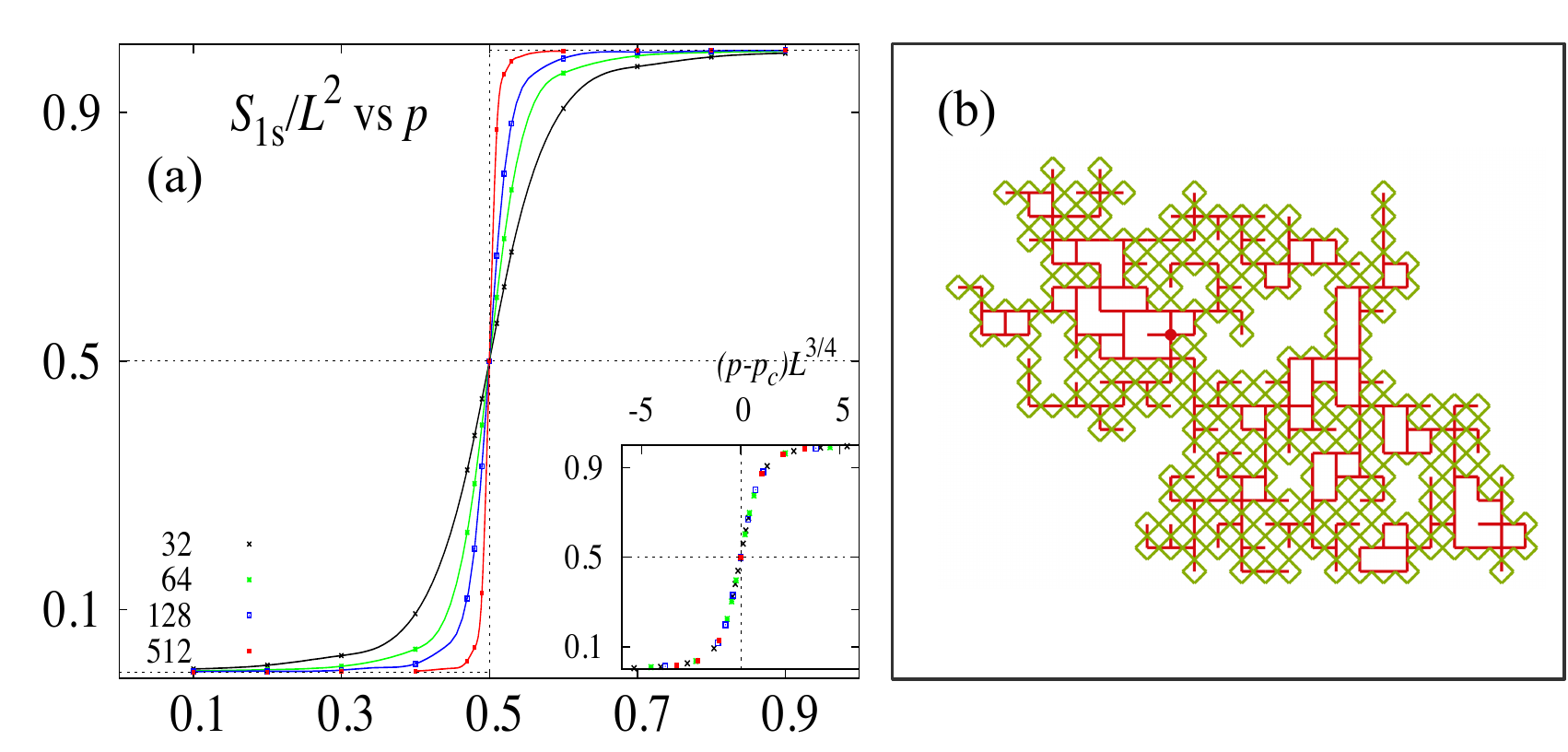}
   \caption{(color online.) (a) The ``solid" size $S_{\rm 1s}/L^2$ of the largest cluster on the triangular lattice vs.\ $p$ for different $L$.   The inset shows the scaling plot  $S_{\rm 1s}/L^2$  versus $(p-p_c)L^{1/\nu}$ with $\nu=4/3$.  When $L \to \infty$ this gives a step function which signifies explosive behavior. 
      (b) An example of a hull walk around a bond percolation cluster of 268 occupied bonds and 239 connected sites.  The 760-step walk (diagonal line segments) connects points on the medial lattice and turns right when encountering an occupied bond and left when encountering a vacant bond, going counterclockwise, and yields the enclosed area or $S_{\rm 1s}$ of 144 square lattice spacings.}
   \label{fig:HullWalkAroundBondPerc}
\end{figure}

Let us consider the ``solid" size $S_{\rm 1s}$ of the largest black cluster with every hole but the largest filled in.
When $p < p_c$,  $S_{\rm 1s}$ will be $O(1)$, while at $p_c$ it  will be $O(L^2)$.
For instance, for s-TR, for $L \rightarrow \infty$ one has $ S_{\rm 1s}/L^2=0$ for $p <p_c$, $1/2$ for $p=p_c$, and $1$ for $p> p_c$, 
as shown by Fig.~\ref{fig:HullWalkAroundBondPerc}(a).
This implies a discontinuity in the ratio $S_{\rm 1s}/L^2$  \cite{SheinmanEtAl15}, which is a signature of  an ``explosive" phenomenon in percolation \cite{AchlioptasDSouzaSpencer09,Ziff09,AraujoAndradeZiffHerrmann11}.  Likewise the size of the largest hole $h_\mathrm{max}/L^2$, which is equivalent to the black solid cluster switching colors,  steps from 1 to 0 at $p_c$.
While the hole problem is based upon standard percolation and its threshold is the same, it is not equivalent to standard percolation since creating the holes requires a global change to the system (identifying the holes and removing all internal sites, and in the NEP case, also identifying the backbone).  It is an explosive feature contained in standard percolation.

\begin{figure}[htbp] 
   \centering
   \includegraphics[width=3.4in]{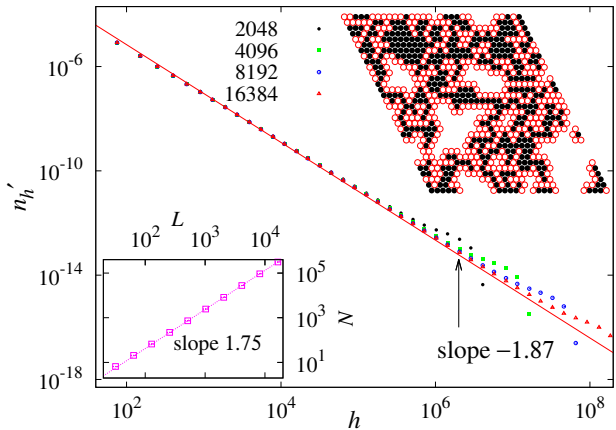}                                                                                                                                                                                                                                                                                                                                                                                                                                                                                                                                                                                                                                                                                                                                                                                                                                                                                                                                                                                                                                                                                                                                                                                                                                                                                                                                                                                                                                                                                                                                         
   \caption{(color online.) Main plot: the rescaled number of holes $n'_h(L,p_c) = n_h L^{d-d_H}$ along the boundary of the largest white cluster broken up by an internal layer of blocked white sites (red circles in inset figure) vs.\ $h$, for systems of different size $L$ (legend).   The slope agrees with the prediction (\ref{eq:taua}).  Inset plot: the total number of boundary holes $N$ as a function of $L$ on a log-log plot, with a slope of $d_H = 7/4$. }
   \label{fig:hull}
\end{figure}

At this transition we have something like a two-phase coexistence (the large black cluster and all its holes except the largest as one phase, and the largest white cluster the other phase).  
The fraction of space filled by the first phase goes from 0 for $p < p_c$ to $1/2$ ($p = p_c$) to 1 ($p>p_c$); see Fig.~\ref{fig:HullWalkAroundBondPerc}(a).  
 The two phase are separated by a fractal boundary with dimension $d_H = 7/4$.  However, there is no phase equilibrium between these two systems and no surface tension between the phases, so in a thermodynamic sense it is not a first-order transition.

We can also envision creating the solid clusters (or holes, switching black and white) by an epidemic or Leath kind of growth process, starting with a seed and adding neighbors with a probability $p$, blocking neighbors for site percolation with probability $1-p$, and filling in the holes. 
  The probability of growing a cluster of $s$ occupied sites is $P_s = s n_s(p)$, where $n_s(p)$ is the number of clusters of size $s$ per site, and the total filled-in area of this cluster will be $A = R^d = s^{d/d_f}$ where $R \sim s^{1/d_f}$ is the radius.  Thus, the mean enclosed area of epidemically grown clusters at $p_c$ will be 
\begin{equation}
\sum_{s=1}^{s_\mathrm{max}} A P_s  \sim \sum_{s=1}^{s_\mathrm{max}}   s^{d/d_f} s^{1-\tau} \sim  \sum_{s=1}^{s_\mathrm{max}} 1 \sim s_\mathrm{max} \sim L^{d_f}
\end{equation}
or, proportional to the mass of the largest cluster.
The quantity $A P_s$ goes to a constant for large $s$ but is not universal: it is 0.04853 for b-SQ but 0.0696 for s-SQ, where in both cases the area is in units of the square of the lattice spacing.   Fig.\ \ref{fig:HullWalkAroundBondPerc}(b) shows a walk around the external hull of a bond cluster, used to find the enclosed area, which gives the size of the filled-in cluster.
Note that the number distribution of enclosed-hull areas has been studied \cite{CardyZiff03};  it is universal and satisfies a Zipf-law form with known constant.  How that distribution relates to the measurements of $A P_s$ is an interesting area for future research.

Finally, we consider another place that holes appear --- when sections of a boundary are broken off.  We consider the largest white hole in the system (site percolation) after all the black sites have been removed.  Next we add a layer of blocked white sites at the internal boundary of hole, as shown in Fig.\ \ref{fig:hull}.  This breaks up the white cluster into many additional holes, some of which can be thought of as fjords into the surrounding black cluster.  We find that the number of holes is proportional to the hull length $L^{d_H}$, and the corresponding $\tau_{h'}$  is correctly predicted:
\begin{equation}
\tau_{h'} = 1 + \frac{d_H}{2} = \frac{15}{8} = 1.875
\label{eq:taua}
\end{equation}
as shown in Fig. \ref{fig:hull}. 
Again, $\tau < 2$.

Note that the appearance of holes is very much a phenomenon of a planar lattice, and these considerations do not apply in higher dimensions.  
On the other hand they should apply to any critical two-dimensional system, such as to the Random Cluster model (a correlated percolation) \cite{FortuinKasteleyn72,Wu82}, 
composed of Fortuin-Kasteleyn clusters for the $q$-state Potts model, where $d_f =2-(6-g)(g-2)/(8g)$ with $q = 2 + 2 \cos (g \pi/2)$ for $ 2 \le g \le 4$ \cite{NienhuisRiedelSchick80,SaleurDuplantier87}, implying $\tau_h = 1 + d_f/2  <2$
for all $0 < q \le 4$.   One can also study the backbones and holes of the random cluster model.  These would be interesting areas of future research.

Thus we have shown that the universality class of the original NEP model is that of holes in backbone percolation, which has an exponent of $\tau_b = 1 + d_B/2 \approx 1.822$.  If we interpret the clusters should instead be equivalent to simple holes, then we have another universality class with $\tau_h = 1 + d_f/2 = 187/96$.   The derived general expression for $\tau$, Eq.\ (\ref{mandelbrot}), is further supported by the case of 
holes cut from the hull, which gives $\tau_{h'} = 1 + d_H/2 = 15/8$.
Likewise, one can conceive of removing the outer layer of the largest backbone hole, which would lead to many additional holes, the number of which 
would be proportional to the hull length $L^{d_{H_b}}=L^{4/3}$ of backbone, and this would yield another case with $\tau_{b'}=1+ d_{H_b}/2=5/3$.
We are not aware if any of these universality classes, all with $\tau < 2$ and all associated with explosive percolation behavior, have been discussed before.

Y. Deng and H. Hu thank the National Natural Science Foundation of China for their support under Grant No.\ 11275185.
Y. Deng acknowledges the Ministry of Education (China) for the Fundamental Research Funds for the Central Universities under Grant No.\ 2340000034.
R. Ziff thanks the University of Science and Technology of China for its hospitality during which this work was written.

\bibliography{bibliography.bib}

\end{document}